\documentclass[preprint,aps,prd,showpacs,nofootinbib,superscriptaddress]{revtex4}

\usepackage{amsfonts}
\usepackage{mathrsfs}
\usepackage{graphicx}
\usepackage{amsmath}
\usepackage{amssymb}
\usepackage{bm}
\usepackage{bbm}
\usepackage{epsfig}
\usepackage{multirow}
\usepackage{float}
\usepackage{array}
\usepackage{tabularx}
\usepackage{xcolor}
\usepackage{enumerate}

\makeatletter
\def\hlinew#1{%
  \noalign{\ifnum0=`}\fi\hrule \@height #1 \futurelet
   \reserved@a\@xhline}
\usepackage{array}
\newcommand{\PreserveBackslash}[1]{\let\temp=\\#1\let\\=\temp}
\newcolumntype{C}[1]{>{\PreserveBackslash\centering}p{#1}}
\newcolumntype{R}[1]{>{\PreserveBackslash\raggedleft}p{#1}}
\newcolumntype{L}[1]{>{\PreserveBackslash\raggedright}p{#1}}
\usepackage{float}

\begin{document}

\title{Lepton number violation in $D$ meson decay}

\author{Hai-Rong Dong\footnote{donghr@ihep.ac.cn}}
\affiliation{Institute of High Energy Physics, Chinese Academy of
Sciences, Beijing 100049, China\vspace{0.2cm}}

\author{Feng Feng\footnote{F.Feng@outlook.com}}
\affiliation{Center for High Energy Physics, Peking University,
Beijing 100871, China\vspace{0.2cm}}

\author{Hai-Bo Li\footnote{lihb@ihep.ac.cn}}
\affiliation{Institute of High Energy Physics, Chinese Academy of
Sciences, Beijing 100049, China\vspace{0.2cm}}

\date{\today}

\begin{abstract}
The lepton number violating process can be induced by introducing a
fourth generation heavy Majorana neutrino, which is coupled to the
charged leptons of Standard Model. There have been many previous
studies on the leptonic number violating decay processes with this
mechanism, we follow the trend to study the process: $D \to K \ell
\ell \pi$ with the same-sign dilepton final states. We restrict
ourself to certain neutrino mass regions, in which the heavy
neutrino could be on shell and the dominant contribution to the
branching fraction comes from the resonance enhanced effect.
Applying the Narrow Width Approximation, we found that upper limit for the branching
fractions for $D^0 \to K^-\ell^+\ell^+\pi^-$ are generally at the order of $10^{-10}$ to $10^{-9}$,
if we take the most stringent upper limit bound currently available in the literature for the mixing matrix elements.
We also provide the constrains, which is competitive compared to the LNV $B$ decays,
on the mixing matrix element $|V_{eN}|^2$
based on the upper limit of $D^0 \to K^- e^+ e^+ \pi^-$ estimated from Monte-Carlo study at BESIII.
\end{abstract}

% 12.38.Bx   Perturbative calculations
% 13.20.Fc [charmed meson] leptonic and semi-leptonic decay
% 13.20.Fs leptonic decay
% 11.30.Fs Baryon number
% 14.60.Pq neutrinos

\maketitle

%%%%%%%%%%%%%%%%%%%%%%%%%%%%%%%%%%%%%%%%%%%%%%%%%%%%%%%%%%%%%%%%%%%%%%%%%%%%%%%%%%%%%%%%%%%%%
\section{Introduction\label{sec1}}

 The discovery of neutrino oscillations~\cite{Fukuda:1998mi,Ahmad:2002jz,Eguchi:2002dm} and observation
 of unexpected large $\theta_{13}$~\cite{An:2012eh} have convincingly shown that
 neutrinos have finite mass and that lepton flavor is violated in
 neutrino propagation. The generation of neutrino masses is still
 one of the fundamental puzzle in particle physics.
 To obtain the non-vanishing mass of neutrinos, one must
make a minimal extension of the Standard Model (SM) by including
three right-handed neutrinos. Once the consensus that the neutrino
is a massive fermion have been reached, another urgent task is to
figure out whether the neutrinos are Dirac or Majorana particles,
the latter case is characterized by being their own antiparticles.

The fact is, at present, the Majorana neutrino is one of the
favorite choices for the most of theories, since the masses of the
observed light neutrinos could be naturally derived from heavy
neutrinos via the so-called ``see-saw'' mechanism
\cite{Georgi:1974sy,Minkowski:1977sc,gellman, yanagida,
glashow,Weinberg:1979sa,Mohapatra:1979ia}. Owing to the new heavy
neutrino¡¯s Majorana nature, it is its own antiparticle, which
allows processes that violate lepton-number conservation by two
units. Consequently, searches for Majorana neutrinos are of
fundamental interest.

There is one promising method to probe the Majorana nature of
neutrinos, i.e., neutrinoless double beta ($0\nu\beta\beta$)
decay\cite{KlapdorKleingrothaus:2001ke,Avignone:2007fu,Doi:1985dx,BiPet87,Bilenky:2001rz,
Petcov:2005yq,Pascoli:2007qh,Bilenky:2012qi,Bilenky:2010zz,Bilenky:2010kd,Mitra:2011qr,Dueck:2011hu,Doi:1992dm}.
As we have mentioned above, the existence of Majorana mass term could induce
the lepton number violation (LNV) by exchange of virtual Majorana
neutrinos between two associated beta decays. Although the first
double-beta decay was proposed as early as 1935 by Goepper-Mayer, it
was until four years later Furry first calculated the
$(0\nu\beta\beta)$ decay based on the Majorana
theory\cite{Furry:1939qr}. These early exploration gave an impetus
to many years of experimental and theoretical research.
It is an interesting question that whether the moderately heavy sterile neutrino with a mass from few hundred MeV to a few GeV exists. If such neutrino exists, the decay rates of these processes can be substantially enhanced by the effect of neutrino-resonance which is induced by taking the mass of virtual Majorana neutrino as a sterile neutrino mass. As a consequence, we may have an chance to observe such LNV processes, or set up stringent upper bounds on sterile neutrino mass and mixing matrix elements.
Thanks to the important significance, these LNV processes have been
extensively studied in both theory~\cite{Atre:2009rg} and
experiment~\cite{Lees:2011hb,BABAR:2012aa,Aaij:2011ex,Aaij:2012zr,Seon:2011ni,Miyazaki:2012mx,Liventsev:2013zz,PDG}.

In recent years, most of the studies of these LNV processes focus on
the three-body and four-body $\Delta L=2$ decays of $K$, $D$ and $B$
mesons, as well as tau lepton decays. The decay processes:
$K^+,D^+,B^+\to \ell^+ \ell^+ M^-$, where $M$ denotes a vector  or
pseudoscalar meson and $\ell=e,\mu,\tau$, have been extensively
studied in \cite{Littenberg:1991ek, Littenberg:2000fg,
Cvetic:2010rw,Zhang:2010um,Ali:2001gsa, Atre:2009rg}. In the
meanwhile, the tau lepton decay $\tau^- \to \ell^+ M_1^- M_2^-$ has
also been discussed in
\cite{Buras:1990fn,Ilakovac:1995wc,Gribanov:2001vv,Helo:2010cw,Cvetic:2010rw}.
Compared to the three-body LNV processes, the studies of four-body
LNV decays are relatively rare. The ${B}\to D\ell\ell\pi$ four-body
decay has been recently calculated in \cite{Quintero:2011yh}, and
the four-body decays $\tau^- \to M^+\ell^- \ell'^- \nu_{\tau}$ are
investigated in \cite{Castro:2012ma}. Recently, more and more
attentions are directed to the four-body LNV decay since these
processes can give more abundant kinematical regions than the
three-body LNV decays, thus we can grope more wide range of the
resonance neutrino mass. In this paper, we will study the four-body
LNV decays $D\to K\pi \ell^+ \ell^+$.

In Particle Data Book (PDG)~\cite{PDG}, the upper limits for the
branching fractions of three-body $D\to \ell^+ \ell^+ M^-$ decays
are at the level of $10^{-6}$ to $10^{-4}$.   For the four-body LNV
$D$ decays, we have less information until recently. In
\cite{Aitala:2000xt}, the branching fractions of $D$ meson decay
$D^0\to M_1^- M_2^- l^+ l^+$ were reported to be the order
$10^{-5}\!\sim \!10^{-4}$.  From experimental side of view, the
BESIII experiment is taking data at open-charm threshold, and an
integrated luminosity of 2.9 fb$^{-1}$ data sample has been
collected at $\psi(3770)$ by the BESIII detector~\cite{Li:2012tr,Li:2006tr}.
%With this data sample,
Estimated from Monte-Carlo (MC) sample of the same luminosity,
the sensitivity of $D\to K \pi \ell^+ \ell^+$ can
reach the level of $10^{-9}$~\cite{qin2013}. It is interesting to
investigate $D\to K\pi \ell^+ \ell^+$ decays mediated by an on-shell
Majorana neutrino.

The paper is organized as follows, in Section~\ref{sec2}, we give a
brief introduction to the theoretical framework involving the LNV
decays; in Section~\ref{sec3}, we sketch some techniques in our
computations, like phase space parametrization and Narrow Width
Approximation, etc.; in Section~\ref{sec4}, we give our numerical
results for the branching fractions of LNV $D$ meson decays; and
finally comes the summary, we also provide the analytical results of
the squared amplitude in the Appendix~\ref{AppendixA}.

%%%%%%%%%%%%%%%%%%%%%%%%%%%%%%%%%%%%%%%%%%%%%%%%%%%%%%%%%%%%%%%%%%%%%%%%%%%%%%%%%%%%%%%%%%%%%
\section{Theoretical Framework\label{sec2}}
We consider the LNV four-body decay of $D$ meson:
\begin{equation}
D(p) \to K(p_1) + \ell^+(p_2) + \ell^+(p_3) + \pi^-(p_4)\;,
\end{equation}
where $D$ with momentum $p$ and $K$ with momentum $p_1$ can be
charged or neutral, two $\ell^+$ are charged leptons with momenta
$p_2$ and $p_3$, respectively, and the charged pion $\pi^-$ has
momentum $p_4$.

Following the previous studies \cite{Atre:2009rg,BarShalom:2006bv}, such
LNV process can be induced through a Majorana neutrino $N$ coupled to the charged leptons $\ell$,
such gauge interactions are described by the following vertex in the Lagrangian:
\begin{equation}
\mathcal{L}=-\frac{\mathit{g}}{\sqrt{2}}W^+_\mu \sum\limits^{\tau}\limits_{\ell=e}V^*_{\ell N}\overline{N^c}\gamma^{\mu}P_L\ell+\mathrm{h.c.},
\end{equation}
where $P_L=\frac{1}{2}(1-\gamma_5)$, $N$ is the mass eigenstate of the fourth generation Majorana neutrino, $V_{\ell N}$ is the mixing matrix between the charged lepton $\ell$ neutrino $\nu_{\ell}$ and heavy Majorana neutrino $N$, their restrictive bounds are reported in \cite{delAguila:2008pw}.
\begin{equation}
{\rm Set\; I:} \quad |V_{eN}|^{2} < 3 \times 10^{-3}, \quad |V_{\mu N}|^{2} < 3 \times 10^{-3}, \quad |V_{\tau N}|^{2} < 6 \times 10^{-3}
\end{equation}

\begin{figure}[t]
\centering
\includegraphics[width=0.9\textwidth]{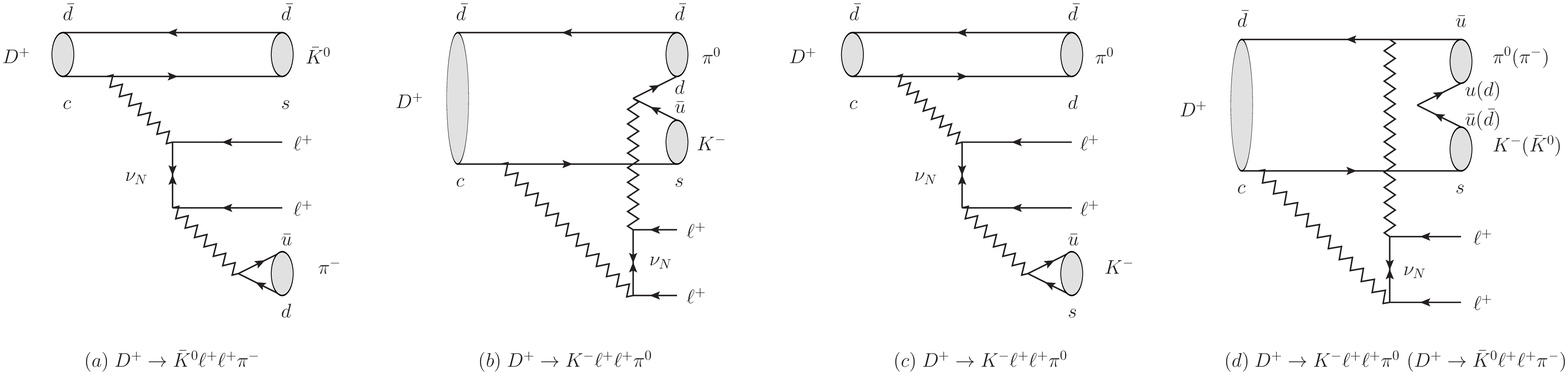}
\includegraphics[width=.75\textwidth]{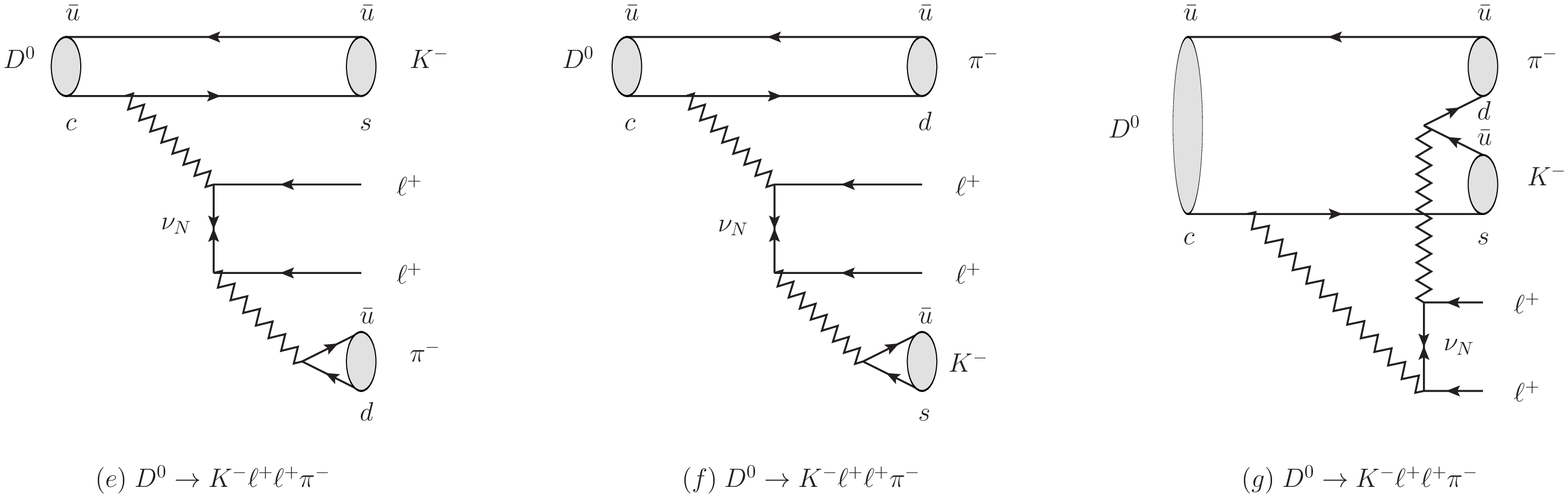}
\caption{ Leading order Feynman diagrams for $\Delta L=2$ charged
and neutral $D$ meson decays.\label{FeynmanDiagram}}
\end{figure}

At leading order, the Feynman diagrams are displayed in
Fig.~\ref{FeynmanDiagram} for the charged and neutral $D$ meson
decays, the diagrams with the charged leptons exchanged are also
included.  If the neutrino mass is from a few hundred MeV to several
GeV, the neutrino propagator in diagrams (a), (c), (e) and (f) in
Fig.~\ref{FeynmanDiagram} could be on-shell, i.e., the neutrino
becomes a resonance, and its contribution to the decay can be much
enhanced due to such neutrino-resonance effect, so the contributions
from other diagrams are negligible compared to such enhanced
diagrams, since the neutrino can not become a resonance in those
diagrams due to the kinematic restrictions. In addition, we note
that diagrams (c) and (f) are suppressed with respect to diagrams
(a) and (e) in Fig.~\ref{FeynmanDiagram} due to smaller
Cabibbo-Kobayashi-Maskawa (CKM) factors
($|V_{cd}V_{us}/V_{cs}V_{ud}| \sim 0.05$). Therefore,
 we only keep diagram (a) and (e) in Fig.~\ref{FeynmanDiagram}
as the dominant contributions.

The transition amplitude for such LNV process can be written as:
\begin{eqnarray}\label{Amplitude}
\mathcal{M} &=& 2 G_F^2 V_{cs} V_{ud} \langle K\vert\bar{s}\gamma^\mu(1-\gamma_5)c\vert D\rangle
\langle \pi|{\bar{u}}\gamma^\nu(1-\gamma_5)d|0\rangle \\
&& \left[V_{\ell N}^2 m_N \bar{u}(p_2)\left(
\frac{\gamma^\mu\gamma^\nu}{q_N^2-m_N^2+i\Gamma_N m_N}
+\frac{\gamma^\nu\gamma^\mu}{q_N'^2-m_{N}^2+i\Gamma_N m_N}\right)
P_R \, v(p_3)\right] \nonumber  \;,
\end{eqnarray}
where $P_R=\frac{1}{2}(1+\gamma_5)$,   $G_F= 1.16639 \times 10^{-5}
\, \mbox{GeV}^{-2}$ is the Fermi constant,  and $V_{cs}$ and
$V_{ud}$ are the CKM matrix element. We have already replaced the
neutrino propagator with its resonant type:
\begin{equation}
\frac{1}{q_N^2-m_N^2} \to \frac{1}{q_N^2-m_N^2+i\Gamma_N m_N} \;,
\end{equation}
and $q_N$ is the momentum of heavy Majorana neutrino, $q'_N$ is the same except with two charged leptons exchanged,
$m_N$ is the mass of such heavy Majorana neutrino, $\Gamma_N$ is the total decay width of the heavy Majorana neutrino.

The decay width of the heavy Majorana neutrino can be obtained by adding up all contributions of neutrino decay channels which can be opened up at the mass $m_N$\cite{Atre:2009rg}:
\begin{equation}
\Gamma_N(m_N) = \sum_s \Gamma(N\to \mbox{final states}) \;
\theta(m_N - \sum_{s \in \mbox{\scriptsize\{final states\}}} m_{s})
\;,
\end{equation}
where $m_s$ in argument of Heaviside $\theta$ function are the masses of final state particles in the corresponding decay channel. All expressions for these decay width can be found in Appendix C of \cite{Atre:2009rg}.

The matrix element involving $\pi$ in Eq.(\ref{Amplitude}) is
related to the decay constant of charged $\pi$ by:
\begin{equation}
 \langle \pi(p_4) |{\bar{u}}\gamma^\nu(1-\gamma_5)d|0\rangle = if_\pi p_4^\nu
 \;,
\end{equation}
where $f_\pi$ is the decay constant of the charged $\pi$.  Then
squared amplitude can be obtained as:
\begin{eqnarray}\label{Suqared Amplitude}
\vert \mathcal M\vert^2 &=&  G_F^4 \left\vert V_{cs}\right\vert^2 \left\vert V_{ud}\right\vert^2
\langle K\vert\bar{s}\gamma^\mu(1-\gamma_5)c\vert D\rangle
\langle K\vert\bar{s}\gamma^{\rho}(1-\gamma_5)c\vert D\rangle^*
\, (i f_\pi p_4^\nu) \, (i f_\pi p_4^{\sigma})^*\\
&&\hspace{-.5cm} {\rm Tr}\left[\big({\rm PN}_2 \gamma_\mu\gamma_\nu + {\rm PN}_3 \gamma_\nu\gamma_\mu\big)
  (1 + \gamma_5)(p\!\!\!/_3 - m_3)(1 - \gamma_5)\big({\rm PN}_2^* \gamma_{\sigma}\gamma_{\rho}
  +{\rm PN}_3^* \gamma_{\mu 1}\gamma_{\nu 1}\big)(p\!\!\!/_2 + m_2)\right]
  \nonumber \;,
\end{eqnarray}
where the factor ${\rm PN}_1$ and ${\rm PN}_2$ are defined as:
\begin{equation}
{\rm PN}_i = \frac{(V_{\ell N})^2 m_N}{(p-p_1-p_i)^2-m_N^2+im_N \Gamma_N} \;.
\end{equation}
The complete analytical expression for the squared amplitude can be found in Appendix \ref{AppendixA}.

%%%%%%%%%%%%%%%%%%%%%%%%%%%%%%%%%%%%%%%%%%%%%%%%%%%%%%%%%%%%%%%%%%%%%%%%%%%%%%%%%%%%%%%%%%%%%
\section{Computation Techniques\label{sec3}}
%====================================================================
\subsection{Kinematics for four-body decay}
\begin{figure}[t]
\includegraphics[width=.6\textwidth]{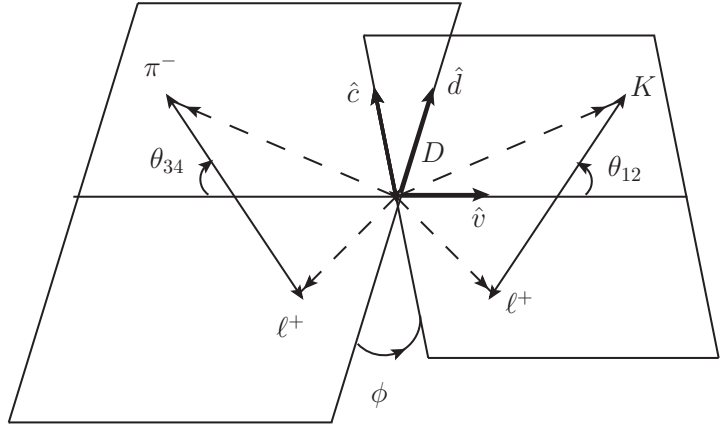}
\caption{Kinematics of four-body decays: $D \to K \ell^+\ell^+ \pi^-$ in the rest frame of $D$ meson.\label{kinematics}}
\end{figure}

The kinematics of the four-body decay $D \to K \ell^+\ell^+ \pi^-$
can be described in terms of five independent variables: $\{s_{12},
s_{34}, \theta_{12}, \theta_{34}, \phi\}$, which have the following
geometrical definitions \cite{delAmoSanchez:2010fd} (see
Fig.~\ref{kinematics}):

\begin{enumerate}[(i)]
\item $s_{12}\equiv (p_1+p_2)^2$ is the square of invariant-mass of $K\ell$ system;
\item $s_{34}\equiv (p_3+p_4)^2$ is the square of invariant-mass of $\pi \ell$ system;
\item $\theta_{12}$ is the angle between the $K$ three-momentum in the $K \ell$ rest frame and the line of flight  of the $K \ell$ in the  $D$ rest frame;
\item $\theta_{34}$ is the angle between the $\pi$ three-momentum in the $\pi \ell$ rest frame and the line of flight  of the $\pi \ell$ in the  $D$ rest frame;
\item $\phi$ is the angle between the normals to the planes defined in the $D$ rest frame by the $K \ell$ pair and the $\pi \ell$ pair.
\end{enumerate}
The angular variables are shown in Fig.~\ref{kinematics}, where
$\bm{K}$ is the $K$ three-momentum in the $K \ell$
center-of-mass(CM) frame and $\bm  \pi$ is the three-momentum of the
$\pi$ in the $\pi \ell$ CM frame. Let $\bm {\hat v}$  be the unit
vector along the $K \ell$ direction in the $D$ rest frame, $\bm
{\hat c}$ the unit vector along the projection of $\bf{K}$
perpendicular to $\bm {\hat v}$, and
  $\bm {\hat d}$ the unit vector along the projection of $\bm {\pi}$ perpendicular
  to $\bm {\hat v}$. We have
\begin{eqnarray}
 \cos \theta_{12}\equiv \frac{{\hat{\bm{v}}} \cdot {\bm K}}{\vert {\bm K} \vert},\,\,\, \cos \theta_{34}\equiv - \frac{{\hat{\bm {v}}} \cdot {\bm \pi}}{\vert{ \bm \pi} \vert},\,\, \cos \phi \equiv {\hat{\bm c}}\cdot {\hat{\bm d}},\,\, \sin \phi \equiv ({\hat{\bm c}}\times {\hat{\bm v}})\cdot {\hat{\bm d}} \;.
 \end{eqnarray}

The Lorentz invariant phase space for the four-body decay is defined as
\begin{eqnarray} \label{Phase Space}
\int  d \Phi_4 &=&  \int \prod_{i=1}^{4} \frac{d^3 p_i}{(2\pi)^3 2 E_i} (2\pi)^4 \delta(p-p_1-p_2-p_3-p_4)\\
&=&\int \frac{d s_{12}}{2\pi} \frac{d s_{34}}{2\pi}d \Phi_2(q_{12},q_{34}) d \Phi_2(p_{1},p_{2}) d \Phi_2(p_{3},p_{4}) \nonumber\\
&=&\int \frac{d s_{12}}{2\pi} \frac{d s_{34}}{2\pi}\left( \frac{\bar \beta}{8\pi} \int \frac{d \cos \theta}{2} \frac{d \phi}{2\pi} \right)\left(\frac{\bar \beta_{12}}{8\pi} \frac{d  \cos \theta_{12}}{2} \frac{d \phi_{12}}{2\pi} \right)\left(\int \frac{\bar \beta_{34}}{8\pi} \frac{d \cos \theta_{34}}{2} \frac{d \phi_{34}}{2\pi} \right) \;,\nonumber
\end{eqnarray}
where $q_{ij} = p_i + p_j$ and $\bar\beta$ and $\bar\beta_{ij}$ are defined as:
\begin{eqnarray}
\bar \beta = \sqrt{1-\frac{2(s_{12}+s_{34})}{s}+\frac{(s_{12}-s_{34})^2}{s^2}}, \,
 \bar \beta_{ij}= \sqrt{1-\frac{2(m_i^2+m_j^2)}{s_{ij}}+\frac{(m_{i}^2-m_{j}^2)^2}{s_{ij}^2}} \,.
\end{eqnarray}
Here we decompose a four-body phase space integral into a product of two-body phase space integrals. This is very useful if one considers a production of two particles, each of which subsequently decays into two-body state.

%====================================================================
\subsection{Narrow Width Approximation}
With the squared amplitude in Eq.(\ref{Suqared Amplitude}) and phase
space parametrization in Eq.(\ref{Phase Space}), we are ready to get
the decay rate for $D \to K \ell \ell \pi$ using the decay rate
formula:
\begin{equation}
\Gamma(D\to K \ell \ell \pi) = \frac{1}{2!} \frac{1}{2m_D} \int d\Phi_4 \vert \mathcal{M} \vert^2
\end{equation}
Let's look at the Eq.(\ref{Amplitude}) again before we start to
perform the phase space integral. In general, $q_N \neq q'_N$, and
it is convenient to split up the individual resonant contributions
 by the Single-Diagram-Enhanced (SDE) multi-channel integration method \cite{Maltoni:2002qb}.
 To do this, we define the functions
\begin{equation}
f_i=\frac{\vert\mathcal{M}_i\vert^2}{\sum_i\vert\mathcal{M}_i\vert^2} \; \vert\sum_i \mathcal{M}_i\vert^2 \,,
\end{equation}
where each $\mathcal{M}_i$ corresponds to the amplitude for a single diagram, then the total amplitude squared is given by
\begin{equation}
\vert\mathcal{M}\vert^2 = \vert\sum_i\mathcal{M}_i\vert^2=\sum_i f_i \,.
\end{equation}
The amplitude squared now splits up into the functions $f_i$ defined
above and the phase space integration can be done for each $f_i$
separately, moreover the peak structure of each $f_i$ is the same as
of the single squared amplitude $\vert \mathcal{M}_i\vert^2$. When
the width $\Gamma_N$ of the heavy Majorana neutrino is very small
compared to the neutrino mass $m_N$, we can apply the Narrow Width Approximation(NAW):
\begin{equation}\label{NAW}
\left.\int\frac{d m_i^2}{(m_i^2-m_N^2)^2+\Gamma_N^2m_N^2}\right\vert_{\Gamma_N\to 0}=\int d m_i^2 \, \delta(m_i^2-m_N^2) \frac{\pi}{\Gamma_N m_N}
\end{equation}
Applying NAW and SDE multi-channel integration method, we can make convenient simplification for the phase space integration and the computation can be carried out in parallel. The contribution from each $f_i$ can be added up after phase space integration.

%%%%%%%%%%%%%%%%%%%%%%%%%%%%%%%%%%%%%%%%%%%%%%%%%%%%%%%%%%%%%%%%%%%%%%%%%%%%%%%%%%%%%%%%%%%%%%%%%%
\section{Numerical Results\label{sec4}}
The matrix element $\langle
K\vert\bar{s}\gamma^\mu(1-\gamma_5)c\vert D\rangle$ for $D$ to $K$
can be parameterized as:
\begin{equation}
\langle K(p_1)\vert\bar{s}\gamma^\mu(1-\gamma_5)c\vert D(p)\rangle =
\left( p_1^\mu+p^\mu - q^\mu \frac{m_D^2-m_K^2}{q^2}\right)f_+(q^2)
+\frac{m_D^2-m_K^2}{q^2} q^\mu f_0(q^2) \;,
\end{equation}
where $q=p-p_1$, and $f_+$ and $f_0$ are two form factors. When we
take the lepton mass as zero, the $f_0(q^2)$ will not contribute,
the strong interaction dynamics can be described by a single form
factor $f_+(q^2)$.
 We use the Modified Pole (MP)~\cite{Becirevic:1999kt} ansatz to parameterize the the form
factor $f_+(q^2)$:
\begin{equation}
f_+(q^2) = \frac{f_+(0)}{\left(1-\frac{q^2}{m_{\rm
pole}^2}\right)\left(1-\alpha_{\rm pole}\frac{q^2}{m_{\rm
pole}^2}\right)} \;,
\end{equation}
where $m_{\rm pole}$ is the pole mass which is predicted to be the
$D^{*-}$ mass, and $\alpha_{\rm pole}$ is a free parameter.  We take the
parameters from CLEO-c measurement~\cite{Besson:2009uv} as follows:
\begin{equation}
f_+(0) = 0.739, \quad m_{\rm pole} = 1.91~{\rm GeV},
\quad\alpha_{\rm pole} = 0.30 \;.
\end{equation}
\vspace{-1cm}
\begin{table}[!ht]
\caption{The masses of mesons and leptons, all the values are taken
from PDG\cite{PDG}.\label{para table1}}
\begin{tabular}
{>{\centering}p{0.15\textwidth}|>{\centering}p{0.1\textwidth}>{\centering}p{0.1\textwidth}>{\centering}p{0.1\textwidth}
>{\centering}p{0.1\textwidth}>{\centering}p{0.1\textwidth}>{\centering}p{0.1\textwidth}>{\hfill}p{0.1\textwidth}<{\hfill\hfill}}
\hlinew{1pt}
meson & $D^+$ & $D^0$ & $K^0$ & $K^-$ & $e^-$ & $\mu^-$ & $\pi^-$ \cr
\hlinew{.1pt}
mass(MeV) & $1869.62$ & $1864.86$ & $497.614$ & $493.677$ & $0.511$ & $105.658$ & $139.57$ \cr
\hlinew{1pt}
\end{tabular}
\end{table}
\begin{table}[!ht]
\caption{Other input parameters used in our case, all the values are taken from PDG\cite{PDG}.\label{para table2}}
\begin{tabular}
{>{\centering}p{0.15\textwidth}|>{\centering}p{0.25\textwidth}>{\centering}p{0.15\textwidth}>{\centering}p{0.15\textwidth}
>{\hfill}p{0.2\textwidth}<{\hfill\hfill}}
\hlinew{1pt}
parameter & $G_F$ & $|V_{ud}|$ & $|V_{cs}|$ & $f_{\pi}$ \cr
\hlinew{.1pt}
value & $1.166\times10^{-5}~ {\rm GeV}^{-2}$ & $0.974$ & $1.006$ & $130.41~{\rm MeV}$ \cr
\hlinew{1pt}
\end{tabular}
\end{table}

We adopt the following values for other input parameters in TABLE~\ref{para table1} and \ref{para table2} in our numerical evaluation.

As for the neutrino mixing matrix elements, we take the most stringent upper
limit currently available in the literature \cite{Atre:2009rg,Helo:2010cw} in our numerical computation, these values are listed in TABLE~\ref{mixing matrix} for both $\vert V_{eN}\vert^2$ and $\vert V_{\mu N}\vert^2$.
\begin{table}[!ht]
\caption{The most stringent upper limit for the neutrino mixing matrix at different majorana neutrino mass $m_N$, the upper limits are extracted from the plots of FIG.3 and FIG.4 in \cite{Atre:2009rg,Helo:2010cw}.\label{mixing matrix}}
\begin{tabular}
{>{\centering}p{0.15\textwidth}|>{\centering}p{0.08\textwidth}>{\centering}p{0.08\textwidth}>{\centering}p{0.08\textwidth}>{\centering}p{0.08\textwidth}>{\centering}p{0.08\textwidth}
>{\centering}p{0.08\textwidth}>{\centering}p{0.08\textwidth}>{\centering}p{0.08\textwidth}>{\hfill}p{0.08\textwidth}<{\hfill\hfill}}
\hlinew{1pt}
$m_N$(MeV) & $150$ & $200$ & $250$ & $300$ & $350$ & $400$ & $450$ & $500$ & $550$ \cr
\hlinew{.1pt}
$\vert V_{eN}\vert^2$($\times10^{-7}$) & $12.0$ & $1.3$ & $1.3$ & $1.3$ & $1.3$ & $1.9$ & $1.9$ & $11.2$ & $9.5$\cr
\hlinew{.1pt}
$\vert V_{\mu N}\vert^2$($\times10^{-7}$) & $19.4$ & $9.3$ & $8.1$ & $2.3$ & $0.8$ & $9.3$ & $6.4$ & $5.1$ & $4.0$\cr
\hline\hline
$m_N$(MeV) & $600$ & $650$ & $700$ & $750$ & $800$ & $850$ & $900$ & $950$ & $1000$ \cr
\hlinew{.1pt}
$\vert V_{eN}\vert^2$($\times10^{-7}$) & $7.5$ & $6.9$ & $6.4$ & $5.4$ & $5.3$ & $4.6$ & $4.3$ & $3.7$ & $3.4$ \cr
\hlinew{.1pt}
$\vert V_{\mu N}\vert^2$($\times10^{-7}$) & $3.5$ & $3.2$ & $3.1$ & $2.5$ & $2.0$ & $1.8$ & $1.7$ & $1.4$ & $1.3$ \cr
\hlinew{1pt}
\end{tabular}
\end{table}

Since $\vert V_{\tau N}\vert^2$ is not controlled by experimental limits due to the absence of experimental data on the LNV or LFV processes involing two $\tau$-leptons, we use an ad hoc assumption
\begin{equation}
\vert V_{eN}\vert^2 \sim \vert V_{\mu N}\vert^2 \sim \vert V_{\tau N}\vert^2 = \vert V_{\ell N}\vert^2
\end{equation}
frequently used in the literature in our computation for total decay width of heavy majorana neutrino $\Gamma_N$. With this assumption, it is easy to derive the following relation between decay rates $\Gamma$ of $D\to K\pi\ell\ell$ with different mixing matrix elements according to the NAW approximation in Eq.~(\ref{NAW}):
\begin{equation}
\frac{\Gamma(m_N, V_{\ell N}(m_N))}{\Gamma(m_N, V'_{\ell N}(m_N))} = \frac{\vert V_{\ell N}(m_N)\vert^2}{\vert V'_{\ell N}(m_N)\vert^2}
\end{equation}

%--------------------------------------
\begin{table}[t]
\caption{\label{Tbl D0}Upper limits on the branching fractions for $D^0 \to K^- +
\ell^+ + \ell^+ + \pi^-$ and $D^0 \to K^- + \mu^+ + \mu^+ + \pi^-$.
The masses of heavy Majorana neutrino are in unit of MeV, and the
total decay width of $D^0$ is $\Gamma_{\rm tot} = 1.605 \times
10^{-9}~{\rm MeV}$.}
\begin{center}
\begin{tabular}
%--------------------------------------
{>{\centering}p{0.07\textwidth}>{\centering}p{0.15\textwidth}>{\centering}p{0.07\textwidth}>{\centering}p{0.15\textwidth}
>{\centering}p{0.005\textwidth}|>{\centering}p{0.005\textwidth}
>{\centering}p{0.07\textwidth}>{\centering}p{0.15\textwidth}>{\centering}p{0.07\textwidth}>{\hfill}p{0.15\textwidth}<{\hfill\hfill}}
\hlinew{1pt}
\multicolumn{5}{c|}{$D^0 \to K^- e^+ e^+ \pi^-$} & \multicolumn{5}{c}{$D^0 \to K^- \mu^+ \mu^+ \pi^-$} \\
\hline $m_N$ & $\Gamma/\Gamma_{\rm tot}$ & $m_N$ &
$\Gamma/\Gamma_{\rm tot}$ &&& $m_N$ & $\Gamma/\Gamma_{\rm tot}$ &
$m_N$ & $\Gamma/\Gamma_{\rm tot}$ \cr \hline $150$ & $1.5\times
10^{-8}$ & $600$ & $2.1\times 10^{-9}$ &&& $150$ & $-$ & $600$ &
$8.3\times 10^{-10}$ \cr \hline $200$ & $2.6\times 10^{-9}$ & $650$
& $1.4\times 10^{-9}$ &&& $200$ & $-$ & $650$ & $5.6\times
10^{-10}$ \cr \hline $250$ & $2.2\times 10^{-9}$ & $700$ &
$9.4\times 10^{-10}$ &&& $250$ & $1.6\times 10^{-9}$ & $700$ &
$3.9\times 10^{-10}$ \cr \hline $300$ & $1.6\times 10^{-9}$ & $750$
& $5.7\times 10^{-10}$ &&& $300$ & $1.3\times 10^{-9}$ & $750$ &
$2.2\times 10^{-10}$ \cr \hline $350$ & $1.3\times 10^{-9}$ & $800$
& $3.6\times 10^{-10}$ &&& $350$ & $4.5\times 10^{-10}$ & $800$ &
$1.2\times 10^{-10}$ \cr \hline $400$ & $1.5\times 10^{-9}$ & $850$
& $2.2\times 10^{-10}$ &&& $400$ & $5.0\times 10^{-9}$ & $850$ &
$6.8\times 10^{-11}$ \cr \hline $450$ & $1.2\times 10^{-9}$ & $900$
& $1.2\times 10^{-10}$ &&& $450$ & $3.0\times 10^{-9}$ & $900$ &
$3.7\times 10^{-11}$ \cr \hline $500$ & $5.4\times 10^{-9}$ & $950$
& $5.6\times 10^{-11}$ &&& $500$ & $2.0\times 10^{-9}$ & $950$ &
$1.7\times 10^{-11}$ \cr \hline $550$ & $3.5\times 10^{-9}$ &
$1000$ & $2.7\times 10^{-11}$ &&& $550$ & $1.3\times 10^{-9}$ &
$1000$ & $7.6\times 10^{-12}$ \cr \hlinew{1pt}
%--------------------------------------
\end{tabular}
\end{center}
\end{table}
%--------------------------------------

%--------------------------------------
\begin{table}[t]
\caption{\label{Tbl D+}Upper limits on the branching fractions for $D^+ \to
\bar{K}^0 + \ell^+ + \ell^+ + \pi^-$ and $D^+ \to \bar{K}^0 + \mu^+
+ \mu^+ + \pi^-$. The masses of heavy Majorana neutrino are in unit
of MeV, and the total decay width of $D^+$ is $\Gamma_{\rm tot} =
6.329 \times 10^{-10}~ {\rm MeV}$.}
\begin{center}
\begin{tabular}
%--------------------------------------
{>{\centering}p{0.07\textwidth}>{\centering}p{0.15\textwidth}>{\centering}p{0.07\textwidth}>{\centering}p{0.15\textwidth}
>{\centering}p{0.005\textwidth}|>{\centering}p{0.005\textwidth}
>{\centering}p{0.07\textwidth}>{\centering}p{0.15\textwidth}>{\centering}p{0.07\textwidth}>{\hfill}p{0.15\textwidth}<{\hfill\hfill}}
\hlinew{1pt}
\multicolumn{5}{c|}{$D^+ \to \bar{K}^0 e^+ e^+ \pi^-$} & \multicolumn{5}{c}{$D^+ \to \bar{K}^0 \mu^+ \mu^+ \pi^-$} \\
\hline $m_N$ & $\Gamma/\Gamma_{\rm tot}$ & $m_N$ &
$\Gamma/\Gamma_{\rm tot}$ &&& $m_N$ & $\Gamma/\Gamma_{\rm tot}$ &
$m_N$ & $\Gamma/\Gamma_{\rm tot}$ \cr \hline $150$ & $2.9\times
10^{-8}$ & $600$ & $5.3\times 10^{-9}$ &&& $150$ & $-$ & $600$ &
$2.1\times 10^{-9}$ \cr \hline $200$ & $6.6\times 10^{-9}$ & $650$
& $3.6\times 10^{-9}$ &&& $200$ & $-$ & $650$ & $1.4\times
10^{-9}$ \cr \hline $250$ & $5.6\times 10^{-9}$ & $700$ &
$2.4\times 10^{-9}$ &&& $250$ & $4.1\times 10^{-9}$ & $700$ &
$1.0\times 10^{-9}$ \cr \hline $300$ & $4.0\times 10^{-9}$ & $750$
& $1.5\times 10^{-9}$ &&& $300$ & $3.3\times 10^{-9}$ & $750$ &
$5.7\times 10^{-10}$ \cr \hline $350$ & $3.1\times 10^{-9}$ & $800$
& $9.3\times 10^{-10}$ &&& $350$ & $1.2\times 10^{-9}$ & $800$ &
$3.2\times 10^{-10}$ \cr \hline $400$ & $3.7\times 10^{-9}$ & $850$
& $5.5\times 10^{-10}$ &&& $400$ & $1.3\times 10^{-8}$ & $850$ &
$1.8\times 10^{-10}$ \cr \hline $450$ & $3.0\times 10^{-9}$ & $900$
& $3.0\times 10^{-10}$ &&& $450$ & $7.6\times 10^{-9}$ & $900$ &
$9.6\times 10^{-11}$ \cr \hline $500$ & $1.4\times 10^{-8}$ & $950$
& $1.4\times 10^{-10}$ &&& $500$ & $5.0\times 10^{-9}$ & $950$ &
$4.3\times 10^{-11}$ \cr \hline $550$ & $8.9\times 10^{-9}$ &
$1000$ & $7.0\times 10^{-11}$ &&& $550$ & $3.0\times 10^{-9}$ &
$1000$ & $2.0\times 10^{-11}$ \cr
\hlinew{1pt}
%--------------------------------------
\end{tabular}
\end{center}
\end{table}
%--------------------------------------
\begin{figure}[!ht]
\includegraphics[width=.7\textwidth]{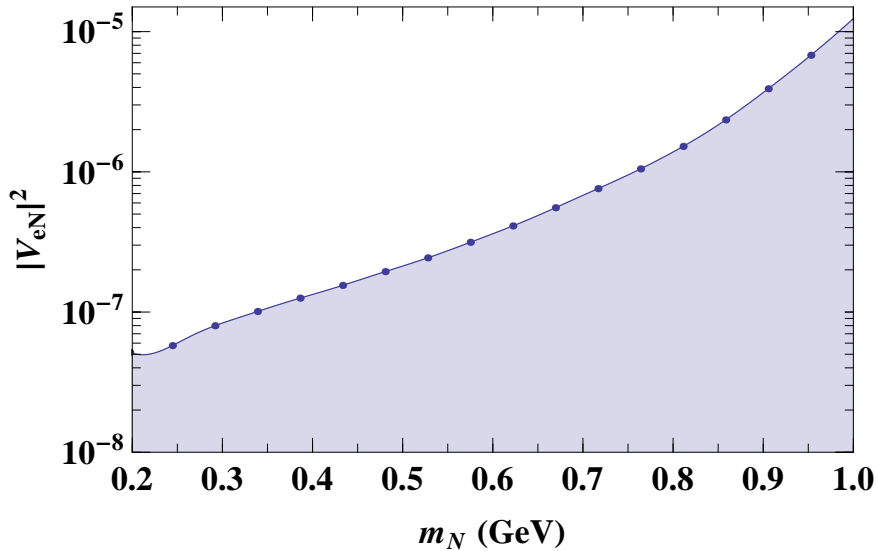}
\caption{Upper limits on $|V_{eN}|^2$ at 90\% confidence level as a
function of the Majorana neutrino mass from the $D^0\to K^- e^+ e^+
\pi^-$ estimated from MC study at BESIII.\label{vvm}}
\end{figure}

To make the heavy mojorana neutrino to be resonant, the neutrino mass $m_N$ should satisfy the following kinematical restrictions:
\begin{equation}
m_\ell + m_{\pi} \le m_N \le m_D - m_\ell - m_K.
\end{equation}
We take the neutrino mass $m_N$ from $150~{\rm MeV}$ to $1000~{\rm
MeV}$, the numerical results for the upper limits of branching fractions are listed
in the TABLE~\ref{Tbl D0} and \ref{Tbl D+}. We find that the
magnitude of the branching fraction for the $D^0\to K^- e^+ e^+
\pi^-$ and $D^0\to K^- \mu^+ \mu^+ \pi^-$ are generally at the order of $10^{-10}$ to $10^{-9}$
with the most stringent mix matrix elements; and the branching fractions for the charged $D$ meson decay
are generally a few times larger than the one for the neutral $D$
meson decay.  We note that upper limits of the branching fractions generally decrease with the neutrino mass increasing.

At the BESIII experiment, estimated from $2.9~{\rm fb}^{-1}$ MC sample, the upper limit for the decay of $D^0\to
K^- e^+ e^+ \pi^-$ obtained in Ref.~\cite{qin2013} is about
$1.0\times 10^{-9}$ at 90\% confidence level. It is already below the upper limit with the most stringent mixing matrix elements available
in the literature for the low majorana neutrino mass $m_N \le 500 {\rm MeV}$. In Fig.~\ref{vvm} we
plot the exclusion regions provided by %the BESIII result. Currently, the BESIII measurement
this MC study. From such estimation, we expect the BESIII experiment can
provide competitive constraints on the
mixing matrix element $|V_{eN}|^2$ compared to the LNV $B$
decays~\cite{Liventsev:2013zz}. In the future, at charm factory,
about 1 ab$^{-1}$ integrated luminosity will be collected per
year~\cite{Asner:2006sd}, and an improvements by three orders of
magnitude on the branching fractions would yield more stringent
constrain on both the mass of Majorana neutrino and the mixing
matrix elements.

\section{Summary}
We have investigated the LNF four-body decay of $D$ meson by
introducing a fourth generation heavy Majorana neutrino $N$, which
is coupled to the charged leptons from the SM, the branching
fractions for such decays are dependent on the heavy neutrino mass
$m_N$, we found that upper limit of the branching fractions are generally at the order of $10^{-10}$ to $10^{-9}$,
if we take the most stringent limit bound currently available in the literature for the mixing matrix element.
We also provide competitive constrains on the mixing matrix element $|V_{eN}|^2$
based on the upper limit of $D^0 \to K^- e^+ e^+ \pi^-$ %from BESIII.
estimated from MC study at BESIII.

\vspace{.5cm}

\vspace{.5cm}
{\bf Note added:} After the calculation was completed and while we were preparing the draft, a related work recently appeared in arXiv \cite{Yuan:2013yba}, which also investigated the lepton-number-violating D meson four-body decay processes. Aside from the different
strategy in parameterizing the $D\to K$ form factor (The authors of \cite{Yuan:2013yba} used the Bethe-Salpeter approach to estimate those form factors, while we use the Modified Pole ansatz whose parameters are directly extracted from the latest published CLEO-c data \cite{Besson:2009uv}), our numerical predictions for the branching ratios are in agreement with theirs in magnitude.

\acknowledgments
We are grateful to Yu Jia for many useful discussions. This research was supported in part by the National Natural Science Foundation of China under
Grants No. 10935012, No. 11125525, the DFG and NSFC (CRC 110), and by the Ministry of Science and Technology of China under Contract No. 2009CB825200 and China Postdoctoral Science Foundation Grant No. 2013M530733.

%\newpage
\appendix
\section{Squared Amplitude\label{AppendixA}}
We give the analytical expression of the squared amplitude in this appendix.
{\footnotesize\begin{eqnarray}
\mathcal{C}_1 &=& f_{\pi}^2 \left|{\rm PN}_2\right|^2 G_F^4 \left| V_{cs}\right|^2 \left| V_{ud}\right|^2 \Big\{ 8 D_0^2 \left(f_0-f_1\right){}^2 m_K^4 \left(-2 m_{\pi }^2 \left(\left(p_{23}+p_{24}-p_{34}\right) m_l^2+m_l^4-2 p_{23} p_{34}\right) \right.\nonumber\\
&&\left. +4 p_{34}^2 \left(m_l^2+p_{23}\right)+m_{\pi }^4 p_{23}\right)+m_K^2 \left(16 f_1 \left(-D_0 \left(f_0-f_1\right) \left(-2 m_{\pi }^2 \left(\left(p_{12}+p_{13}+p_{23}+p_{24}-p_{34}\right) m_l^2 \right.\right.\right.\right.\nonumber\\
&&\left.\left.\left.\left.+m_l^4-p_{14} p_{23}+p_{13} p_{24}-p_{12} p_{34}-2 p_{23} p_{34}\right)+4 p_{34} \left(\left(p_{14}+p_{34}\right) m_l^2+p_{14} p_{23}-p_{13} p_{24}+p_{12} p_{34} \right.\right.\right.\right.\nonumber\\
&&\left.\left.\left.\left. +p_{23} p_{34}\right)+m_{\pi }^4 p_{23}\right)+2 f_1 m_{\pi }^2 p_{23}-4 f_1 p_{24} p_{34}\right)-16 D_0^2 \left(f_0-f_1\right){}^2 m_D^2 \left(-2 m_{\pi }^2 \left(\left(p_{23}+p_{24}-p_{34}\right) m_l^2 \right.\right.\right.\nonumber\\
&&\left.\left.\left.+m_l^4-2 p_{23} p_{34}\right)+4 p_{34}^2 \left(m_l^2+p_{23}\right)+m_{\pi }^4 p_{23}\right)\right)+8 D_0^2 \left(f_0-f_1\right){}^2 m_D^4 \left(-2 m_{\pi }^2 \left(\left(p_{23}+p_{24}-p_{34}\right) m_l^2 \right.\right.\nonumber\\
&&\left.\left.+m_l^4-2 p_{23} p_{34}\right)+4 p_{34}^2 \left(m_l^2+p_{23}\right)+m_{\pi }^4 p_{23}\right)+16 D_0 \left(f_0-f_1\right) f_1 m_D^2 \left(-2 m_{\pi }^2 \left(\left(p_{12}+p_{13}+p_{23} \right.\right.\right.\nonumber\\
&&\left.\left.\left.+p_{24}-p_{34}\right) m_l^2+m_l^4-p_{14} p_{23}+p_{13} p_{24}-p_{12} p_{34}-2 p_{23} p_{34}\right)+4 p_{34} \left(\left(p_{14}+p_{34}\right) m_l^2+p_{14} p_{23}\right.\right.\nonumber\\
&&\left.\left.-p_{13} p_{24}+p_{12} p_{34}+p_{23} p_{34}\right)+m_{\pi }^4 p_{23}\right)+8 f_1^2 \left(-2 m_{\pi }^2 \left(\left(2 p_{12}+2 p_{13}+p_{23}+p_{24}-p_{34}\right) m_l^2+m_l^4\right.\right.\nonumber\\
&&\left.\left.+4 p_{12} p_{13}-2 p_{14} p_{23}+2 p_{13} p_{24}-2 p_{12} p_{34}-2 p_{23} p_{34}\right)+4 p_{34} \left(\left(2 p_{14}+p_{34}\right) m_l^2+4 p_{12} p_{14}+2 p_{14} p_{23}\right.\right.\nonumber\\
&&\left.\left.-2 p_{13} p_{24}+2 p_{12} p_{34}+p_{23} p_{34}\right)+m_{\pi }^4 p_{23}\right) \Big\}
\end{eqnarray}}\vspace{-.8cm}
{\footnotesize\begin{eqnarray}
\mathcal{C}_2 &=& f_{\pi }^2 \left| {\rm PN}_3\right|^2 G_F^4 \left| V_{cs}\right|^2 \left| V_{ud}\right|^2 \Big\{ 8 D_0^2 \left(f_0-f_1\right){}^2 m_K^4 \left(-2 m_{\pi }^2 \left(\left(p_{23}-p_{24}+p_{34}\right) m_l^2+m_l^4-2 p_{23} p_{24}\right) \right.\nonumber\\
&&\left.+4 p_{24}^2 \left(m_l^2+p_{23}\right)+m_{\pi }^4 p_{23}\right)+m_K^2 \left(16 f_1 \left(-D_0 \left(f_0-f_1\right) \left(-2 m_{\pi }^2 \left(\left(p_{12}+p_{13}+p_{23}-p_{24}+p_{34}\right) m_l^2\right.\right.\right.\right.\nonumber\\
&&\left.\left.\left.\left.+m_l^4-p_{14} p_{23}-p_{13} p_{24}-2 p_{23} p_{24}+p_{12} p_{34}\right)+4 p_{24} \left(\left(p_{14}+p_{24}\right) m_l^2+p_{14} p_{23}+p_{13} p_{24}+p_{23} p_{24}\right.\right.\right.\right.\nonumber\\
&&\left.\left.\left.\left.-p_{12} p_{34}\right)+m_{\pi }^4 p_{23}\right)+2 f_1 m_{\pi }^2 p_{23}-4 f_1 p_{24} p_{34}\right)-16 D_0^2 \left(f_0-f_1\right){}^2 m_D^2 \left(-2 m_{\pi }^2 \left(\left(p_{23}-p_{24}+p_{34}\right) m_l^2\right.\right.\right.\nonumber\\
&&\left.\left.\left.+m_l^4-2 p_{23} p_{24}\right)+4 p_{24}^2 \left(m_l^2+p_{23}\right)+m_{\pi }^4 p_{23}\right)\right)+8 D_0^2 \left(f_0-f_1\right){}^2 m_D^4 \left(-2 m_{\pi }^2 \left(\left(p_{23}-p_{24}+p_{34}\right) m_l^2\right.\right.\nonumber\\
&&\left.\left.+m_l^4-2 p_{23} p_{24}\right)+4 p_{24}^2 \left(m_l^2+p_{23}\right)+m_{\pi }^4 p_{23}\right)+16 D_0 \left(f_0-f_1\right) f_1 m_D^2 \left(-2 m_{\pi }^2 \left(\left(p_{12}+p_{13}+p_{23}\right.\right.\right.\nonumber\\
&&\left.\left.\left.-p_{24}+p_{34}\right) m_l^2+m_l^4-p_{14} p_{23}-p_{13} p_{24}-2 p_{23} p_{24}+p_{12} p_{34}\right)+4 p_{24} \left(\left(p_{14}+p_{24}\right) m_l^2+p_{14} p_{23}\right.\right.\nonumber\\
&&\left.\left.+p_{13} p_{24}+p_{23} p_{24}-p_{12} p_{34}\right)+m_{\pi }^4 p_{23}\right)+8 f_1^2 \left(-2 m_{\pi }^2 \left(\left(2 p_{12}+2 p_{13}+p_{23}-p_{24}+p_{34}\right) m_l^2+m_l^4\right.\right.\nonumber\\
&&\left.\left.-2 \left(p_{14} p_{23}+\left(p_{13}+p_{23}\right) p_{24}\right)+2 p_{12} \left(2 p_{13}+p_{34}\right)\right)+4 p_{24} \left(\left(2 p_{14}+p_{24}\right) m_l^2+2 p_{14} p_{23}+p_{23} p_{24}\right.\right.\nonumber\\
&&\left.\left.+2 p_{13} \left(2 p_{14}+p_{24}\right)-2 p_{12} p_{34}\right)+m_{\pi }^4 p_{23}\right) \Big\}
\end{eqnarray}}\vspace{-.8cm}
{\footnotesize\begin{eqnarray}
\mathcal{C}_3 &=& f_{\pi }^2 {\rm PN}_2^* {\rm PN}_3 G_F^4 \left| V_{cs}\right|^2 \left| V_{ud}\right|^2  \Big\{
8 D_0^2 \left(f_0-f_1\right){}^2 \left(p_{23} m_{\pi }^4+2 \left(m_l^4+p_{23} m_l^2+p_{23} \left(p_{24}+p_{34}\right)\right) m_{\pi }^2\right.\nonumber\\
&&\left.-2 \left(m_l^2 \left(p_{24}^2+p_{34}^2\right)-2 p_{23} p_{24} p_{34}\right)\right) m_D^4+16 D_0 \left(f_0-f_1\right) f_1 \left(p_{23} m_{\pi }^4+2 \left(m_l^4+\left(p_{12}+p_{13}+p_{23}\right) m_l^2\right.\right.\nonumber\\
&&\left.\left.+p_{23} \left(p_{14}+p_{24}+p_{34}\right)\right) m_{\pi }^2-2 \left(\left(p_{24}^2+p_{34}^2+p_{14} \left(p_{24}+p_{34}\right)\right) m_l^2+p_{13} p_{24}^2+p_{12} p_{34}^2-p_{12} p_{24} p_{34}\right.\right.\nonumber\\
&&\left.\left.-p_{13} p_{24} p_{34}-2 p_{23} p_{24} p_{34}-p_{14} p_{23} \left(p_{24}+p_{34}\right)\right)\right) m_D^2+\zeta \left(-32 i \left(m_{\pi }^2+2 p_{14}+p_{24}+p_{34}\right) f_1^2\right.\nonumber\\
&&\left.-32 i D_0 \left(f_0-f_1\right) m_D^2 \left(m_{\pi }^2+p_{24}+p_{34}\right) f_1+32 i D_0 \left(f_0-f_1\right) m_K^2 \left(m_{\pi }^2+p_{24}+p_{34}\right) f_1\right)\nonumber\\
&&+8 D_0^2 \left(f_0-f_1\right){}^2 m_K^4 \left(p_{23} m_{\pi }^4+2 \left(m_l^4+p_{23} m_l^2+p_{23} \left(p_{24}+p_{34}\right)\right) m_{\pi }^2-2 \left(m_l^2 \left(p_{24}^2+p_{34}^2\right)-2 p_{23} p_{24} p_{34}\right)\right)\nonumber\\
&&+8 f_1^2 \left(p_{23} m_{\pi }^4+2 \left(m_l^4+\left(2 p_{12}+2 p_{13}+p_{23}\right) m_l^2+4 p_{12} p_{13}+p_{23} \left(2 p_{14}+p_{24}+p_{34}\right)\right) m_{\pi }^2\right.\nonumber\\
&&\left.-2 \left(\left(p_{24}^2+p_{34}^2\right) m_l^2-4 p_{14}^2 p_{23}+2 p_{14} \left(\left(p_{24}+p_{34}\right) m_l^2+2 p_{13} p_{24}-p_{23} p_{24}+2 p_{12} p_{34}-p_{23} p_{34}\right)\right.\right.\nonumber\\
&&\left.\left.-2 \left(p_{13} \left(p_{34}-p_{24}\right) p_{24}+p_{34} \left(p_{12} p_{24}+p_{23} p_{24}-p_{12} p_{34}\right)\right)\right)\right)+m_K^2 \left(-16 D_0^2 \left(f_0-f_1\right){}^2 \left(p_{23} m_{\pi }^4\right.\right.\nonumber\\
&&\left.\left.+2 \left(m_l^4+p_{23} m_l^2+p_{23} \left(p_{24}+p_{34}\right)\right) m_{\pi }^2-2 \left(m_l^2 \left(p_{24}^2+p_{34}^2\right)-2 p_{23} p_{24} p_{34}\right)\right) m_D^2\right.\nonumber\\
&&\left.-16 D_0 f_1 \left(f_0 \left(p_{23} m_{\pi }^4+2 \left(m_l^4+\left(p_{12}+p_{13}+p_{23}\right) m_l^2+p_{23} \left(p_{14}+p_{24}+p_{34}\right)\right) m_{\pi }^2\right.\right.\right.\nonumber\\
&&\left.\left.\left.-2 \left(\left(p_{24}^2+p_{34}^2+p_{14} \left(p_{24}+p_{34}\right)\right) m_l^2+p_{13} p_{24}^2+p_{12} p_{34}^2-p_{12} p_{24} p_{34}-p_{13} p_{24} p_{34}-2 p_{23} p_{24} p_{34}\right.\right.\right.\right.\nonumber\\
&&\left.\left.\left.\left.-p_{14} p_{23} \left(p_{24}+p_{34}\right)\right)\right)+f_1 \left(p_{23} m_{\pi }^4-2 \left(m_l^4+\left(p_{12}+p_{13}-p_{23}\right) m_l^2-2 p_{23}^2+p_{14} p_{23}-p_{23} p_{24}\right.\right.\right.\right.\nonumber\\
&&\left.\left.\left.\left.-p_{23} p_{34}+2 p_{24} p_{34}\right) m_{\pi }^2+2 \left(\left(p_{24}^2-4 p_{34} p_{24}+p_{34}^2+p_{14} \left(p_{24}+p_{34}\right)\right) m_l^2+p_{13} p_{24}^2+p_{12} p_{34}^2-4 p_{24} p_{34}^2\right.\right.\right.\right.\nonumber\\
&&\left.\left.\left.\left.-4 p_{24}^2 p_{34}-p_{12} p_{24} p_{34}-p_{13} p_{24} p_{34}-6 p_{23} p_{24} p_{34}-p_{14} p_{23} \left(p_{24}+p_{34}\right)\right)\right)\right)\right)
 \Big\} + {\rm h.c.}
\end{eqnarray}}

where $p_{ij} = p_i \cdot p_j$, $\zeta = \varepsilon^{\mu\nu\rho\sigma} p_{1\mu} p_{2\nu} p_{3\rho} p_{4\sigma}$ (we adopt the convention $\varepsilon^{0123}=1$ for the Levi-Civita tensor), and $D_0$ are defined as:
\begin{eqnarray}
D_0 &=& \frac{1}{m_\pi^2+2\left( m_K^2+p_{23}+p_{24}+p_{34} \right)}
\end{eqnarray}

and the amplitude squared is
\begin{equation}
\vert \mathcal{M} \vert^2 = \mathcal{C}_1 + \mathcal{C}_2 + \mathcal{C}_3 \;.
\end{equation}

\end{document}